\documentclass{article}
\usepackage{ijcai20}
\usepackage{comment}
\usepackage{times}
\usepackage{enumitem}

\usepackage{soul}
\usepackage{url}
\usepackage[hidelinks]{hyperref}
\usepackage[utf8]{inputenc}
\usepackage[small]{caption}
\usepackage{graphicx}
\usepackage{amsmath}
\usepackage{booktabs}
\usepackage{csquotes}
\usepackage[T1]{fontenc}
\usepackage[utf8]{inputenc}
\usepackage{natbib}

\usepackage{amssymb}
\usepackage{pifont}
\usepackage{tikz}
\usepackage{multirow}
\usepackage{xcolor}

\usepackage[textsize=footnotesize]{todonotes}
\definecolor{Gray}{gray}{0.95}

\title{Clustering of Social Media Messages for Humanitarian Aid Response during Crisis}

\author{
Swati Padhee$^1$\footnote{Research work was done while Swati
was interning at Dataminr Inc. and presented at the AI for Social Good, Harvard CRCS Workshop 2020 (https://aiforgood2020.github.io).}\and
Tanay Kumar Saha$^2$\and
Joel Tetreault$^2$\And
Alejandro Jaimes$^2$\\
\affiliations
$^1$Wright State University, Dayton, OH\\
$^2$Dataminr Inc., New York, NY\\
\emails
padhee.2@wright.edu, \{tsaha, jtetreault, ajaimes\}@dataminr.com
}

\begin{document}

\maketitle

\begin{abstract}
Social media has quickly grown into an essential tool for people to communicate and express their needs during crisis events. Prior work in analyzing social media data for crisis management has focused primarily on automatically identifying actionable (or, informative) crisis-related messages. In this work, we show that recent advances in Deep Learning and Natural Language Processing outperform prior approaches for the task of classifying informativeness and encourage the field to adopt them for their research or even deployment. We also extend these methods to two sub-tasks of informativeness and find that the Deep Learning methods are effective here as well.

\end{abstract}

\section{Introduction}

\label{sec:Introduction}

During large-scale disasters, humanitarian organizations seek timely and reliable information to understand the overall impact of the crisis in order to respond effectively. Social media platforms, such as Twitter, have emerged as an essential source of timely, on-the-ground information for first responders~\citep{vieweg2014integrating, castillo2016big}. Public Information Officers utilize online social media to gather actionable crisis-related information and provide it to the respective humanitarian response organizations~\citep{castillo2016big}.  

Manually monitoring and classifying millions of incoming social media messages during a crisis is not possible in a short time frame as most organizations lack the resources and workforce. Thus, it is necessary to leverage {\it automatic} methods to identify informative messages to assist humanitarian organizations and crisis managers. An example of an informative message on Twitter during Hurricane Dorian is:  ``\textit{All Tidewater Dental Locations are collecting Donations for the Victims of Hurricane Dorian! Items needed: (PLEASE SHARE) Non-perishable foods, Bug Repellent, Blankets, Clean clothing, Socks, Wipes, Toiletries}". To date, most works in this area have focused on leveraging classical machine learning methods~\citep{imran2014aidr,caragea2016identifying,nguyen2016rapid} for detecting informative messages at a reasonable accuracy. 

Our paper makes two contributions. First, we show that recent Deep Learning and Natural Language Processing methods, specifically state-of-the-art pre-trained language models such as BERT (Bidirectional Encoder Representations from Transformers)~\citep{devlin2018bert} and RoBERTa~\citep{liu2019roberta}, yield substantial improvements in this task and encourage the field to adopt them. We also analyze how robust these methods are to new crisis events.

\begin{figure*}
  \includegraphics[width=\textwidth,scale=0.70]{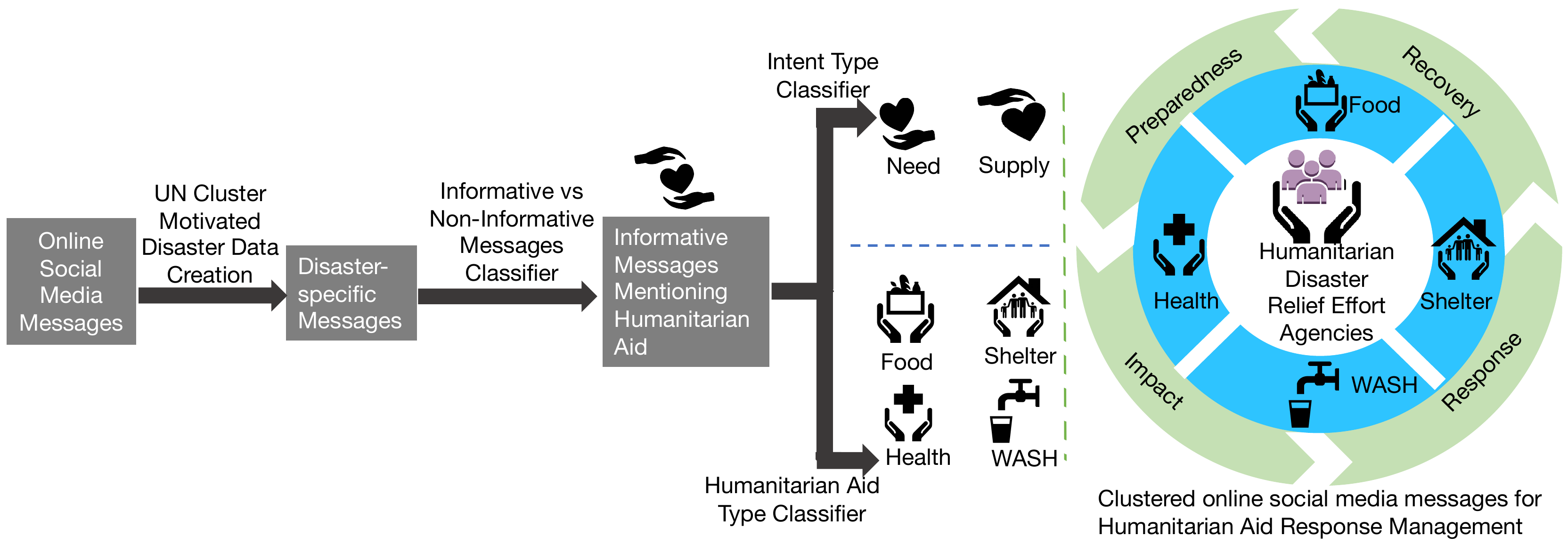}
  \caption{Overview of the UN Cluster motivated Humanitarian Aid Response Management}
   \label{overview}
\end{figure*}

Second, prior work has tended to focus on the binary classification task of whether a message is informative or not. However, more refined classifications can be of further assistance. We propose two new classification tasks motivated by the need of humanitarian organizations to cluster actionable crisis-related messages into their {\it intent} (``Need" and or ``Supply"), and {\it humanitarian aid type} (``Food", ``Shelter", ``Health", and ``WASH"), where WASH stands for (Water, Sanitation, and Hygiene). The latter task is based on the UN Humanitarian Reform process, which outlines eleven primary needs or clusters that humanitarian organizations should track during an emergency. Figure \ref{overview} presents an overview of our proposed architecture for the UN cluster motivated humanitarian aid response management.  

In short, we show that recent neural approaches offer substantial gains across three humanitarian aid classification tasks, of which two are new. The analyses above can be applied to any social media platform, but for this research study, we use messages posted on Twitter during Hurricane Dorian, a Category 5 hurricane that impacted North America in 2019. In Section \ref{sec:Methodology}, we detail the process for constructing a dataset for each of the three tasks. In Section \ref{experimentalDetails}, we describe the models, experimental setup, and results. Finally, in Section \ref{discussion}, we discuss how well the top model performs on unseen crisis events and conclude with future directions in Section \ref{sec:Conclusion}. 

We provide our keywords\footnote{https://github.com/swatipadhee/Crisis-Aid-Terms.git} and annotation guidelines to the community to make our study reproducible. 

\section{Related Work}
\label{sec:RelatedWork}
The task of detecting informative messages from Twitter, or other social media, is typically treated as a three-step process. First, simple filters are employed to extract tweets relevant to the event. Second, machine learning methods are developed to extract informative tweets from the first step automatically. A third step may follow in which the informative tweets are labeled with more specific distinctions of informativeness. We detail the related work for these steps below.

Extracting disaster-specific social media messages usually relies on two types of keyword matching: (a) {\it lexicons}, or keywords related to the disaster, to extract tweets mentioning disaster keywords \citep{purohit2014emergency,hodas2015disentangling,imran2013practical}, and (b) {\it location terms} to extract all tweets associated with the areas impacted by the disaster~\citep{mahmud2012tweet,waqascampfiremissing}. The lexicon-based approach introduces noise (low precision), while the location-based approach has shallow coverage (low recall) and might fail to capture a considerable fraction of relevant tweets. We use both lexicon and location-specific terms in collecting tweets from the Hurricane Dorian that hit in 2019.

Prior works have filtered actionable, informative tweets from large datasets during disasters \citep{nguyen2015tsum4act,caragea2016identifying,zhang2016semi} employing classical machine learning as well as deep learning-based techniques for the classification task~\citep{madichetty2019detecting,imran2014aidr,caragea2016identifying,zhang2016semi,caragea2016identifying,neppalli2018deep,nguyen2016rapid}. More recently, deep learning methods have been successful for a host of NLP tasks such as summarization~\citep{wu2018learning}, machine translation~\citep{edunov2018understanding}, Named entity recognition~\citep{baevski2019cloze} etc. The amount of work leveraging deep learning methods for emergency response tasks is limited. Of note are \cite{jain2019estimating}, which experiments with embeddings such as BERT \cite{devlin2018bert}, ELMo \cite{peters2018deep}, GloVe \cite{pennington2014glove} and word2vec \cite{mikolov2013efficient}, and \cite{alam2020standardizing} which experiments with CNNs and BERT. Informativeness classification varies based on the dataset, but it is possible to achieve performance as high as 0.87 F-score \cite{alam2020standardizing}. In our work, we experiment with a range of machine learning and deep learning methods, including a recent one, RoBERTa~\citep{liu2019roberta}.

Finally, there is some work into further classifying informative messages from the prior stage. Examples of categories include caution or advice, information source, people, casualties, damage, and donations \citep{imran2013practical,neppalli2018deep,alam2020standardizing,madichetty2019detecting,maas2019facebook}\footnote{For a more detailed summary of the different categories and datasets, we refer the reader to \citep{alam2020standardizing}, Section 8.2}. The Multilingual Disaster Response Dataset\footnote{https://appen.com/datasets/combined-disaster-response-data/} covers $36$ humanitarian categories, including Food, Shelter, Water, Clothing, Medical Help, and Medical Products. The humanitarian categorizations of this dataset do not directly align with the need of UN cluster-specific humanitarian organizations, and also, the annotations are not granular enough to differentiate between ``Need" and ``Supply". Similar to informativeness classification, classical machine learning, and deep learning algorithms \citep{alam2020standardizing,jain2019estimating} are mostly used for humanitarian task type classification. We use similar methods for our tasks and show that a recent state-of-the-art language model, RoBERTa performs the best in informativeness classification as well as the newly proposed tasks.

\section{Data Creation}
\label{sec:Methodology}

\subsection{Unsupervised Data Extraction for Hurricane Dorian}
\label{sec:data}

\begin{table*}
\centering
\resizebox{0.80\linewidth}{!}{%
\begin{tabular}{l||cc||c|c||c|c|c|c}
\toprule
\multirow{4}{*}{\textbf{Models}} & \multicolumn{2}{c||}{Informativeness} & \multicolumn{2}{c||}{Intent Type} &  \multicolumn{4}{c}{Aid Type} \\
& \multicolumn{2}{c||}{Classification}   & \multicolumn{2}{c||}{Classification}  &  \multicolumn{4}{c}{Classification} \\
 & \multicolumn{2}{c||} {Total: $1,208$} & \multicolumn{2}{c||} {Total: $1,010$} &   \multicolumn{4}{c} {Total: $1,010$} \\
& \multicolumn{2}{c||} {\textbf{Train:} $966$; \textbf{Test:} $242$} & \multicolumn{2}{c||} {\textbf{Train:} $808$; \textbf{Test:} $202$} &   \multicolumn{4}{c} {\textbf{Train:} $808$; \textbf{Test:} $202$} \\

\cline {2-9}
\cline{2-9}
&   & & & & & \\
&    &  	& {\textbf{Need}} & \textbf{Supply} & {\textbf{Food}}		    &  \multicolumn{1}{c|}{\textbf{Shelter}} 	 &  \multicolumn{1}{c|}{\textbf{Health}}  &  \multicolumn{1}{c}{\textbf{WASH}}\\
 &  Acc (\%) & F1 (\%) & F1 (\%) & F1 (\%) & F1 (\%)  & F1 (\%)  & F1 (\%)  & F1 (\%)\\
\toprule
MNB    & $83.67$ & $83.32$   & $57.69$   & $62.67$   &$77.74$             &$60.25$          &$74.14$             &$65.42$\\
LR       & $77.67$ & $77.78$ & $62.87$  &  $67.38$    &$88.59$            &63.32          &$74.15$            &$64.39$\\
BERT      & $85.67$ & $87.81$ & $67.57$   & $65.39$     &$74.74$             &$63.19$          &$74.83$            &$80.77$ \\
RoBERTa    & {$\mathbf{87.67}$} & {$\mathbf{89.20}$} & {$\mathbf{71.98}$} & $\mathbf{75.98}$ &{$\mathbf{89.53}$}   &{$\mathbf{69.02}$}  &{$\mathbf{78.89}$}   &{$\mathbf{85.81}$}\\
\bottomrule

\end{tabular}%
}

\caption{Categorization Results of Social Media Messages for Humanitarian Aid Response}
\label{res:t4}
\end{table*}

We utilize humanitarian help-type keywords and disaster-specific location terms to extract tweets posted during Hurricane Dorian from Aug 24, 2019, to Sep 23, 2019. We design custom queries by combining the generic disaster-specific keywords used in previous work \citep{crisismmd2018icwsm,firoj2018twitter,olteanu2014crisislex} along with: (a) UN cluster-based lexical keywords released in \citep{temnikova2015emterms}, (b) humanitarian aid type-specific keywords \citep{niles2019social}, and (c) generic disaster location terms (e.g., ``Bahamas"). We utilize a publicly available Python library (GetOldTweets3\footnote{https://pypi.org/project/GetOldTweets3/}) with those custom queries to extract relevant tweets during Hurricane Dorian. This process results in  37,768 unique tweets that serve as our dataset for results and discussion. We remove all user information from the tweets (i.e., handles).

\subsection{Labeled Dataset Creation for Three Humanitarian Tasks}

\label{sec:amt}

Next, we use the tweets from above to build three labeled datasets for three respective supervised tasks: Informativeness, Intent Type, and Aid Type. For each dataset, we use the Amazon Mechanical Turk (AMT) platform\footnote{https://www.mturk.com} to generate ground-truth labels, which we compare system predictions to. Annotation instructions for each task can be found in Table~\ref{annotation}.

\noindent\textbf{Task 1: Informativeness Classification:} 

We sample a set of $1,208$ tweets from the collected $37,768$ unique tweets (we remove tweets with cosine similarity higher than $0.85$) uniformly at random and employ paid expert workers from AMT to generate ground-truth labels of informativeness (whether the tweet is informative or not). Among 1,208 tweets, 482 (39.91\%) were labeled as Informative.

\noindent\textbf{Tasks 2 \& 3: Intent Type and Aid Type Classification:} 

Based on the labeled dataset for Task 1, we develop a binary classification model (please see Section ~\ref{results} for detailed results) and use the RoBERTa model to predict labels on the remaining tweets. The model predicted $14,073$ ($37.26$\%) out of 37,768 tweets as informative. For Tasks 2 and 3, we sample $1,010$ tweets uniformly at random from those 14k informative tweets. For each task and tweet, we got labels from $5$ AMT ``master level" annotators. We decide the final labels for each tweet based on the agreement of three or more annotators. To measure the quality of the annotations, the authors manually annotated 300 tweets and observed a substantial agreement with the majority label produced by the AMT annotators (Cohen's Kappa score~\citep{cohen1960coefficient}
 of $0.71$). We focus on the following humanitarian aid types: food, shelter (temporary or permanent home, basic living needs like clothes or electricity, etc.), water, sanitation, hygiene, and health support.
For Task 2, AMT workers labeled $781$ ($77.33$\%) tweets as ``Need"   and $470$ ($46.54$\%) as ``Supply". As expected, ``Need" tweets are more prevalent than ``Supply" tweets. For Task 3, AMT workers labeled $470$ ($46.54$\%) as ``Food", $291$ ($28.81$\%) as ``Shelter", $164$ ($16.24$ \%) as ``Health" and $276$ ($27.33$ \%) as ``WASH".

\begin{table*}[htbp]
\centering
\smallskip
\scalebox{0.75}{

\begin{tabular}{l|l}
\toprule
\textbf{Task} & \textbf{Instructions} \\
\midrule

Task 1: Informativeness & Given a tweet, select “Yes" if the tweet is talking about either people requesting humanitarian help during a hurricane \\
&  or that help is on the way. Humanitarian help includes food,
shelter, water, hygiene, mental, or physical health support. \\
& Select “No" if the tweet is not talking about any of the humanitarian help types.\\ \hline
Task 2: Intent Type & Select ``Need" if a tweet contains a mention of the need for humanitarian help, regardless of who mentions it and for whom. \\
& Select ``Supply" if a tweet contains a mention of the supply of humanitarian help. Select ``Both" if, in a tweet, there is mention \\ 
& of both need and supply.  If the tweet is NOT about either need or supply, please select ``None of the above." \\ \hline
Task 3: Aid Type: & If there is only one help type, pick one. If there are multiple help types, pick all of them that are relevant. If a tweet is NOT \\
& about any of the help types, please select “None of the above." The choices are: “Food," “Shelter," “Health,"
“Water, Sanitation, \\ 
& and Hygiene (WASH),“ or “None of the
above." \\

\bottomrule
\end{tabular}
}
\caption{Annotation Instructions for Mechanical Turk}
\label{annotation}
\end{table*}

\begin{table*}[ht]
\centering
\smallskip
\scalebox{0.80}{

\begin{tabular}{l|l|c}
\toprule
\textbf{Tweet texts} & \textbf{Human Label} &\textbf{RoBERTa} \\
\midrule
A Tamil-English translator needed. \#FloodSL \#SriLanka & Non-Informative  & Non-Informative\\
SFHS English Department has you covered with your back to school, post Harvey, supply needs! Stop by D103! & Informative & Informative \\
Meet Irma!! The go-to LulaRoe top. It is loose, knit high-lo tunic with fitted sleeves!\#lularoe \#lularoeirma & Informative & Non-Informative \\

\bottomrule
\end{tabular}
}
\caption{Informativeness Classification on Other Crisis Events Datasets}
\label{informativeQual}
\end{table*}

\section{Experiments}
\label{experimentalDetails}

With the three datasets in place, we can benchmark different modeling approaches head to head.  For each dataset, we remove URLs, image links, numbers, hashtags, mentions, non-ASCII characters from tweets, and contract multiple spaces into a single space. 
We use Micro-F1 for binary classification task ({Task 1}), and Macro-F1 for multi-label multi-class tasks ({Task 2} and {Task 3}). We also report accuracy for the informativeness classification task.  We split each dataset into train ($80$\%) and test ($20$\%) partitions. We run our experiments five times and report the average value for each metric.

\subsection{Models Compared}
We present our tasks' performance on two baseline traditional machine learning algorithms using TF-IDF embeddings and a linear softmax classifier using two contextual language model embeddings. 

\noindent\textbf{Multinomial Na\"ive Bayes (MNB)} is a learning technique built upon the theory that the features representing the data points are conditionally independent of each other for a given class. We use TF-IDF based embeddings of documents as features for this model.

\noindent\textbf{LR} is designed using a linear classification function 
that predicts the probability of a data belonging to a particular class. We use TF-IDF based embeddings of documents as features for this model. We use the Scikit-learn pipeline to generate TF-IDF vector representations\footnote{https://scikit\-learn.org/}.

\noindent\textbf{BERT} \citep{devlin2018bert} is a document representation learning model which looks into both left and right context of a word to learn the representation. We choose the pre-trained BERT-base model and add a task-specific fine-tuning layer on top of the BERT architecture for the classification tasks.

\noindent\textbf{RoBERTa} \citep{liu2019roberta} is a language model similar to BERT, but, trained by modifying the design strategies in BERT to achieve better performance in downstream tasks. We add a task-specific fine-tuning layer on top of the RoBERTa architecture for the classification task.

\subsection{Results}
\label{results}
\noindent\textbf{Task 1 Results:}
Table \ref{res:t4} (First part) reports the results for our first classification task (informativeness vs non-informativeness). In this task, RoBERTa outperforms all other models. The model achieves an accuracy of $87.67$\%, and an F1 score of $89.20$\%. The closest competitor of RoBERTa is BERT, which is also based on contextual embeddings. These results indicate that contextual embeddings have much better discriminative power than the traditional TF-IDF based embeddings. 

\noindent\textbf{Task 2 Results:} Table \ref{res:t4} (Second Part) reports the results for the second humanitarian task of classifying a tweet as either Need or Supply. Again, RoBERTa performs best with a Micro-F1 measure of $71.98$\% for predicting ``Need" and that of $75.98$\% for predicting ``Supply." 
Similar to Task 1, BERT is the second best model. 

\noindent\textbf{Task 3 Results:}  Table \ref{res:t4} (Third Part) shows the results for the classification of UN cluster-based humanitarian categories. RoBERTa achieves the highest accuracy and Micro-F1 across all four categories by a wide margin. 

We then use RoBERTa to predict labels on all of the $14,073$ informative tweets filtered using the classifier trained in Task 1. Although our proposed Tasks 2 and 3 are fine intent-driven categorization of humanitarian aid types, we observe results with respect to BERT in the same spirit of ~\cite{jain2019estimating} where they reported BERT performing poorer than Word2Vec or Glove embeddings in information-type classification. Out of the $14,073$ informative tweets, RoBERTa predicted $8,370$ ($59.48$\%) tweets seeking for help and $9,296$ ($66.1$\%) tweets supplying help. As a given tweet can mention both need and supply messages, the model predicted both the labels for $3,593$ tweets. On the same $14,073$ informative tweets, RoBERTa model predicted Food as a label for $7,940$ ($56.42$\%) tweets, Shelter as a label for $3,543$ ($25.2$\%),  2359 ($16.76$\%) as Health, and $4,362$ ($30.1$\%) as WASH. The percentage of tweets predicted correctly reflects the significance of tweets to be channeled to the respective humanitarian organizations. 

\subsection{Discussion}
\label{discussion}
We analyze how a model trained on one crisis event could generalize to unseen crisis events. In an ideal case, such a model would perform well across different crisis events, and no additional data collection and retraining is necessary to deploy. In our case, we investigate how well an Informative classifier trained on Hurricane Dorian can perform on other crisis datasets for the Informativeness task. We use a publicly available dataset of tweets collected during Hurricane Harvey, Hurricane Maria, and Sri Lanka floods~\citep{crisismmd2018icwsm}. RoBERTa achieved an accuracy of $82.12$\% in ``Sri Lanka Floods", but performed poorly on ``Hurricane Harvey" and ``Hurricane Maria" (achieved $54.30\%$ and $59.85\%$ accuracy, respectively). These low results indicate that we need a better strategy for domain adaptation and also require adjustment of labels. In Table \ref{informativeQual}, we list a set of predictions using our best informativeness classifier (RoBERTa) on other crisis datasets~\citep{crisismmd2018icwsm}. Interestingly, RoBERTa classified the third tweet (third row) in Table \ref{informativeQual} as ``Non-Informative" contrary to the human-provided label. Nonetheless, the predicted label is correct based on our task definition (Section \ref{sec:amt}) because the tweet is not talking about any humanitarian help and thus ``Non-Informative".

\section{Conclusion \& Future Work}
\label{sec:Conclusion}
In this work, we show that recent neural approaches offer substantial gains across three humanitarian aid classification tasks. We introduce two additional levels of abstraction (UN Cluster motivated clustering) on top of informativeness classification and show with high accuracy that these tasks can be automated, which can be beneficial to the humanitarian organizations in assessing, prioritizing, and mobilizing the needs of the affected community. Our results show that specifically state-of-the-art language models such as RoBERTa~\citep{liu2019roberta}, yield substantial improvements in these tasks, and we encourage the field to adopt them. In the future, we plan for a qualitative evaluation of our tasks by showing the clustered tweets to the personnel from the respective UN clusters.   

\bibliographystyle{named}
\bibliography{ijcai20}

\end{document}